\documentclass[a4paper]{article}

\usepackage{INTERSPEECH2019}
\usepackage{lipsum}

\usepackage[hidelinks]{hyperref}

\title{Generative Adversarial Network based Speaker Adaptation for High Fidelity WaveNet Vocoder}
\name{Qiao Tian$^1$, Xucheng Wan$^2$, Shan Liu$^1$}

\address{Cloud and Smart Industries Group, Tencent Technology Co., Ltd}
\email{$^1${\{briantian,shiningliu\}}@tencent.com, $^2$xucheng.wan@wisc.edu$^*$}

\begin{document}

%
\maketitle
\newcommand\blfootnote[1]{%
\begingroup
\renewcommand\thefootnote{}\footnote{#1}%
\addtocounter{footnote}{-1}%
\endgroup}

\begin{abstract}
\label{sec:abs}
Although state-of-the-art parallel WaveNet has addressed the issue of real-time waveform generation, there remains problems. Firstly, due to the noisy input signal of the model, there is still a gap between the quality of generated and natural waveforms. Secondly, a parallel WaveNet is trained under a distillation framework, which makes it tedious to adapt a well trained model to a new speaker. To address these two problems, in this paper we propose an end-to-end adaptation method based on the generative adversarial network (GAN), which can reduce the computational cost for the training of new speaker adaptation. Our subjective experiments shows that the proposed training method can further reduce the quality gap between generated and natural waveforms. 

\end{abstract}

\blfootnote{*This paper was done during Xucheng Wan's internship in CSIG, Tencent Technology Co., Ltd}


\section{Introduction}
\label{sec:intro}

In recent years, deep learning has made great progress in the field of speech synthesis. The state-of-the-art approach Tacotron2 \cite{wang2017tacotron}, which proposes an end-to-end acoustic model with modified WaveNet as neural vocoder\cite{oord2016wavenet}, is able to produce high fidelity synthesized audio. 
Compared with the conventional statistical parametric speech synthesis methods \cite{zen2015unidirectional} combining with long short term memory (LSTM) and traditional vocoders \cite{morise2016world, kawahara1999restructuring}, this approach makes the synthesized speech greatly closer towards natural speech in both speech quality and prosody.

Parellel WaveNet \cite{oord2017parallel} is proposed for real-time generation of speech based on the original WaveNet. It alleviates the enormous computational burden of the original auto-regressive  WaveNet while preserving its relative high performance. Core idea of this parallel WaveNet that we employ for our system is the inverse autoregressive flows(IAFs) where sampling process is performed in parallel so that the inference can be implemented much faster than real-time. However, there are still two issues that remained to be addressed in practical applications.
Firstly, training the entire system can be super slow since the basic training procedure is still auto-regressive. The training pipeline of the parallel WaveNet is relatively tedious since it is trained following the model distillation framework \cite{hinton2015distilling}. Under such framework, a well learned auto-regressive  WaveNet is required as the teacher model to guide the training of the student model which is our target parallel WaveNet.
And training both teacher and student models could take weeks. 
In addition, sufficient data are required to train teacher model for the new speaker.
Secondly, although generated speech is of good quality, there still exists a gap between generated and natural speech. 
This is due to the noisy input signal of the parallel WaveNet model, which results in lots of detailed information missing in high frequency domain of the generated speech.

In this paper, we propose an adaptation framework to adapt a well learned parallel WaveNet to a speaker with merely few hours of training data.
We replaced the distillation component in training framework with a generative adversarial component \cite{goodfellow2014generative}. 
The minimax training trick of generative adversarial network (GAN) makes the generated samples undistinguishable from real samples. The discriminator of the GAN can capture some subtle differences between generated waveforms and natural audios, which are usually neglected by the auto-regressive  teacher WaveNet, and it helps the generator to produce audios of higher fidelity.
The contribution of this paper includes: 1) We propose an end-to-end speaker adaptation for high fidelity neural vocoder based on GAN. The training of the proposed framework is much more efficient than the original distillation framework, such as parallel WaveNet and Clarinet \cite{ping2018clarinet}.
2) We use the GAN to further reduce the gap between generated and natural speech. 

This paper is organized as follows: 
In Section \ref{sec:gan}, we will briefly review the basic background of GAN.
Then the proposed method will be given in Section \ref{sec:agan}.
Experimental details and results will be given in  Section \ref{sec:exps}.
Lastly in Section \ref{sec:cons}, conclusions and potential future research directions are presented.


\section{Generative Adversarial Network}
\label{sec:gan}

\begin{figure}[t]
	\centering
	\includegraphics[width=7cm,height=3cm]{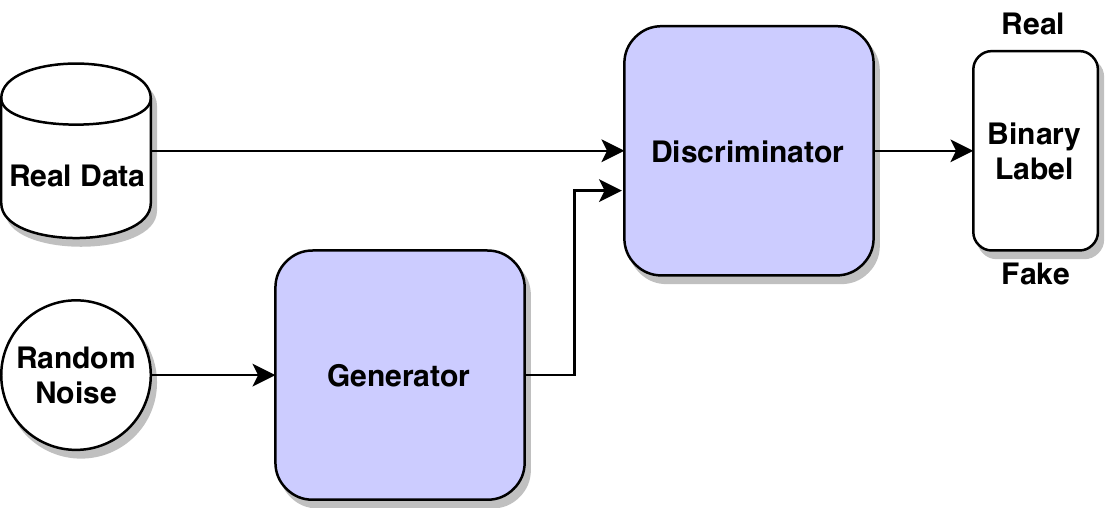}
	\caption{The architecture of generative adversarial network}
	\label{fig:gan}
\end{figure}

Generative Adversarial Network is a new framework proposed in recent years, which has been proven to be able to generate impressive samples in the field of computer vision.
As shown in Fig.~\ref{fig:gan}, a typical GAN model consists of two sub-networks: a Discriminator network (D) and a Generator network (G). 
The generator network learns to map a simple distribution $p_{\bm{z}}(\bm{z})$ to a complex distribution $P_g(\bm{x})$, where $\bm{z}$ denotes the random noise sample and $\bm{x}$ denotes the target data sample. 
The generator is trained to make the generated sample distribution $P_g(\bm{x})$ undistinguishable from real data distribution $P_d(\bm{x})$.
On the contrary, the discriminator is trained to identify the generated (fake) samples against data (real) samples which makes the adversarial training a minimax game.
For conditional sample generation tasks, such as speech synthesis, an additional condition vector $\bm{c}$ is usually added to the input of both the generator and discriminator, which yields the conditional GAN (cGAN) model \cite{odena2016conditional}.
The training objective of the cGAN is formulated as
\begin{align} 
\label{acgan}
  \min\limits_{G} & \max\limits_{D} ~ V(D, G) = E_{\bm{x}, \bm{c} 
\sim p_{d}(\bm{x}, \bm{c})} \left[\log D(\bm{x}, \bm{c}) \right] \notag \\
    &+ E_{\bm{z} \sim p_{\bm{z}}(\bm{z}), c \sim p_{D}(\bm{c})} \left[\log (1 - D(G(\bm{z}, \bm{c}), \bm{c})\right].
\end{align}


The minimax training process of the original GAN is unstable and difficult to converge. And it usually results in mode collapse problem where the samples from the input distribution all map to the same output which the discriminator can not distinguish from the real data.
A lot of tricks are therefore proposed to improve the training and ensure model to learn realistic distribution.
In order to alleviate mode collapse and also address the problem of vanishing gradients caused by sigmoid cross-entropy loss, the least-squares GAN (LSGAN) \cite{mao2017least},
is proposed by replacing the cross-entropy loss with a least-squares binary coding loss. 
The training objectives for the discriminator and generator of the LSGAN are defined as followings:
\begin{align} 
\label{lsgan_d}
  \min\limits_{D} ~ V_{\mathrm{LSGAN}}&(D) = \frac {1}{2} E_{\bm{x}, \bm{c} \sim p_{d}(\bm{x}, \bm{c})}\left[(D(\bm{x}, \bm{c}) - 1)^2 \right] \notag \\
    &+ \frac {1} {2} E_{\bm{z} \sim p_{\bm{z}}(\bm{z}), \bm{c} \sim p_{d}(\bm{c})} \left[D(G(\bm{z}, \bm{c}), \bm{c})^2\right]
\end{align}
\begin{align} 
\label{lsgan_g}
  \min\limits_{G} ~ &V_{\mathrm{LSGAN}}(G) = \notag \\ 
  &\frac {1} {2} E_{\bm{z} \sim p_{\bm{z}}(\bm{z}), \bm{c} \sim p_{d}(\bm{c})}\left[(D(G(\bm{z}, \bm{c}), \bm{c}) - 1)^2 \right]
\end{align}

The LSGAN has been applied to speech enhancement (SEGAN) \cite{pascual2017segan} which generates clean speech signal conditioning on noisy speech signal.
An additional $L_{1}$ norm loss is used in learning the parameters of the G network of SEGAN, enabling it to benefit from adversarial training to product much cleaner speech waveform.
This $L_{1}$ norm based loss term for the generator is defined as follows:
\begin{align} 
\label{segan}
  \min\limits_{G} ~ &V_{\mathrm{SEGAN}}(G) = \lambda ||G(\bm{z}, \widetilde {\bm{x}}) - \bm{x}||_1 + \notag \\
  & \frac {1}{2} E_{z \sim p_{\bm{z}}(\bm{z}), \widetilde {\bm{x}} \sim p_{d}(\widetilde{\bm{x}})} [((D(G(\bm{z}, \widetilde {\bm{x}}), \widetilde {\bm{x}}) - 1)^2],
\end{align}
where $\tilde{\bm{x}}$ denotes the input noisy signal and a hyper parameter $\lambda$ is used to balance the GAN loss and $L_1$ loss.

\section{WaveNet Adaptation Using GAN}
\label{sec:agan}

The original auto-regressive WaveNet is a model that can generate perfect speech waveform. 
Different from the conventional vocoders, such as STRAIGHT \cite{kawahara1999restructuring} and WORLD \cite{morise2016world}, the WaveNet vocoder doesn't depend on the source-filter assumption of speech signal. 
This makes it a perfect vocoder that can avoid the problems of excitation extraction.
However, due to its auto-regressive nature, the waveform generation is unbearably slow (100 times slower than real time or more on a Nvidia Tesla P40 GPU). 


\subsection{Parallel WaveNet Vocoder}

The parallel WaveNet addressed the inference problem by using the inverse auto-regressive  flow (IAF) \cite{kingma2016improved}, which can perform 30 times faster than real time on a Nvidia Tesla P40 GPU. 
However its training is very hard and tricky.

IAF is a method that enables the model to convert the input noise signal into speech waveform. Using noise signal as inputs allows the model to compute in parallel which is key to real-time generation. 
However, IAF is difficult to optimize directly because of the requirement of auto-regressively computed log-likelihood loss. 

\cite{oord2017parallel} proposed a probability density distillation method to distill the student WaveNet efficiently from auto-regressive WaveNet with mixture of logistic (MoL) output distribution \cite{salimans2017pixelcnn++}. 
Therefore, the student WaveNet can generate audios whose fidelities are close to that of the auto-regressive WaveNet. 
On the other hand, the training of the model starts from training a time-consuming teacher auto-regressive WaveNet.

\subsection{WaveNet Adaptation}

Speaker adaptation is a commonly adopted method for fast building of acoustic models for speech synthesis and speech recognition, especially for cases where training data are limited.
If applying speaker adaptation directly to the parallel WaveNet model, 
we will need to apply it to both teacher and student models.
This would make the adaptation training of a new speaker extremely slow and tedious.

\begin{figure}[t]
   \centering
   \includegraphics[width=8cm,height=5.6cm]{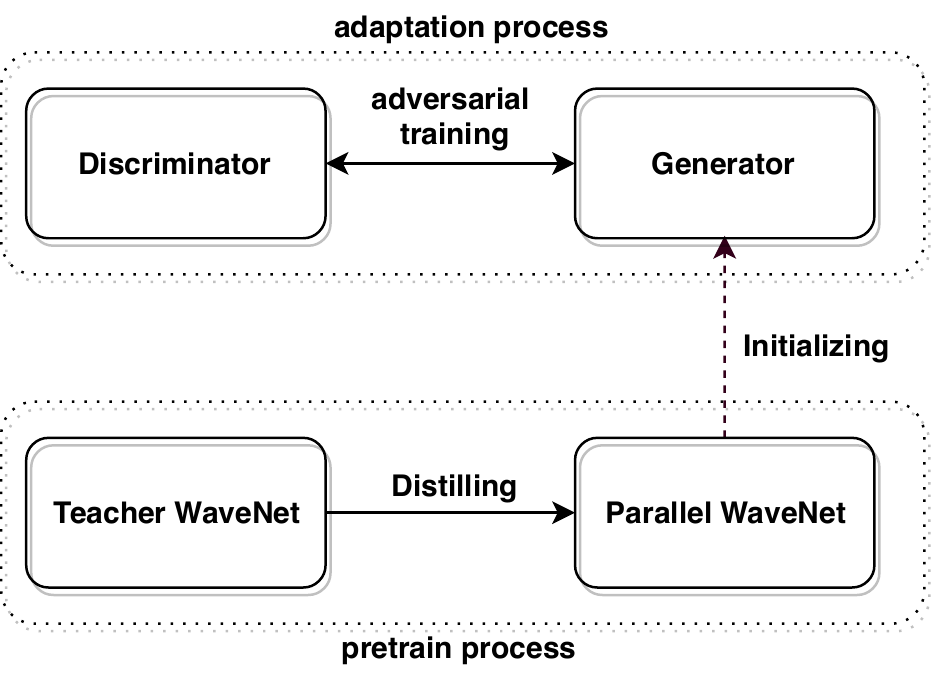}
\caption{The architecture of speaker adaptation of parallel WaveNet using GAN.}
\label{fig:agan_arc}
\end{figure}

Therefore, in this paper we propose to employ the GAN framework to accelerate  this adaptation training process and thus improve the efficiency of training the entire parallel WaveNet system. 
As showed in Fig.~\ref{fig:agan_arc}, we adapt parallel WaveNet to a new speaker based on the adversarial training method by replacing the distilling teacher model with a discriminator from a GAN.

Specifically for the adaptation GAN (AGAN), a parallel WaveNet of one speaker was pre-trained in advance. In this pre-training process, we apply the same model structure as proposed in \cite{oord2017parallel}. But instead of using the same power-loss in the original parallel WaveNet system, we adopt a new loss term - the mean square loss of log-scale STFT-magnitude (log-mag loss) \cite{arik2018fast} which we will explain in detail shortly.

At the adaptation phase, the pre-trained model is used to initialize the parameters of the generator of the GAN. We apply the least-square loss \cite{mao2017least} to stablize the training of adaptation. However, it is difficult for a vocoder trained only using a single adversarial loss to produce speech waveform with high fidelity. Because least-square loss term only learns the information in time domain while the detailed information in frequency domain is neglected. This idea can also be demonstrated by the experiment result of training the model with a single Kullback-Leibler (KL) loss in the original parallel WaveNet\cite{oord2017parallel}. 
We therefore use an additional log-mag loss which has been proven to be effective in capturing spectral details during the training process.
The log-mag loss is computed in frequency domain and it is defined as following:
\begin{equation} 
\label{log-mag}
\begin{split}
L_{\mathrm{log-mag}}(\bm{x}, \bm{x}^\prime) &= ||\log (|\mathrm{STFT}(\bm{x})| + \epsilon) \\
&- \log (|\mathrm{STFT}(\bm{x}^\prime)| + \epsilon)||_1
\end{split}
\end{equation}
where $L_{1}$ norm is used and $\epsilon$ is a very small value to ensure the positivity of spectral magnitude.

We construct our discriminator network using a non-causal dilated convolution structure\cite{rethage2018wavenet}, similar to the architecture of a non-autoregressive  WaveNet, to identify the generated (fake) waveform against the recorded (real) waveform. For this discriminator, we build the network with 10 dilated convolution layers without sacrificing discrimination performance at sample scale. And for our adaptation model, the mel-domain spectrograms are used as conditional input, which is represented by $c$ in this paper.

In detail, for each sentence, we sample the waveform $\bm{x}'$ from the output distribution of the generator network.
Then $\bm{x}'$ is fed to the discriminator network to evaluate the D loss against samples from real data distribution. 
The loss of the generator is defined as 
\begin{align} 
\label{sagan}
  \min\limits_{G} V_{\mathrm{AGAN}}(G) = L_{\mathrm{log-mag}}&(G(\bm{z}, \bm{c}), \bm{x}) + \notag \\
  \frac {\lambda} {2} E_{\bm{z} \sim p_{\bm{z}}(\bm{z}), \bm{c} \sim p_{d}(\bm{c})} &[((D(G(\bm{z}, \bm{c}), \bm{c}) - 1)^2].
\end{align}
where $z$ denotes the input random noise， $c$ denotes the mel-spectrum and $p$'s represent the sample distributions accordingly.

\section{Experiments}
\label{sec:exps}

\subsection{Data Set}
\label{ssec:data}

We use two different datasets for the two training phases of our experiment.
The initial parallel WaveNet model was pre-trained on our internal speech dataset, which contains 12 hours of mandarin speech records by a female speaker.
For the adaptation phase, we use the public LJSpeech dataset \cite{ljspeech17} to evaluate the performance of the proposed speaker adaptation GAN. 
The audio for pre-training phase is re-sampled at 24 kHz, while in the adaptation phase, the original 22.05 kHz sampling rate of LJSpeech is preserved.
LJSpeech dataset contains 13,100 short audio clips of public domain English speech data from a speaker. 
The lengths of audio clips range from 1 to 10 seconds and the total length is approximately 24 hours. 
We randomly select 2000 audio clips, which add up to about 3 hours, as our training data.

It is worth noting that we selected Mandarin speech sampled at 24 kHz to train the basic model while in the adaptation phase we used English speech sampled at 22.05 kHz. Mandarin is a tone-based language and English speech replies more on phonemes.  By conducting our experiments on different languages and at different sampling rates, it is further demonstrated that our model with proposed adaptation GAN method is powerful enough to handle most training cases.

 
\subsection{Model setup}
\label{ssec:ourgan}

Following the configuration of acoustic analysis in Tacotron2 \cite{shen2018natural}, we extracted 80-dim mel-spectrograms as the local acoustic condition for neural vocoders with a frame shift of 256 points and frame length of 2048 points. 
The initial parallel WaveNet was trained with 1500k steps with a teacher MoL WaveNet trained on the same dataset.
In adaptation phase, we adopted the Adam optimizer \cite{kingma2014adam} for the AGAN. 
The noam scheme \cite{vaswani2018tensor2tensor} for learning rate was used with 4k warm-up step. 
The AGAN model was trained with batch size of 4 clips and max sample length 24000.
For comparison, another parallel WaveNet was adapted by distilling using data of the target speaker.

Both generator and discriminator in our GAN structure use adam optimizer. The discriminator of AGAN is trained with a random initialization. 
Its architecture is a non-causal WaveNet with 10 dilated convolution layers using filter size 3 and max dilation rate of 10. 
We add Leaky ReLU \cite{maas2013rectifier} activation function with $\alpha=0.2$ after each layer of convolution except the last output layer. 
The discriminator also uses mel-spectrograms as local condition which is up-sampled to sample scale by a 4-layer de-convolution network.
The learning rate of the generator and discriminator were set to 0.005 and 0.001 respectively. 
In the first 50k steps, we freeze the parameters of the generator in order to better learn a discriminator. 
Then for the next 100k steps till model converges, the generator and discriminator are adversarial trained simultaneously. 
We find that the coefficient $\lambda$ of adversarial loss can, to some extend, reflect the fidelity of the generated waveform. 
We achieve a relatively good result by setting $\lambda = 1.5$ and another experiment with $\lambda=0.05$ is set for comparison.


\subsection{Experimental results}
\label{ssec:results of audio}

We adopt the commonly used Mean Opinion Score (MOS) to subjectively evaluate our proposed GAN-based speaker adaptation framework. 
In order to ensure that the results are convincing enough, we randomly select 30 sentences that are not included in the training set from the dataset.
Three models are compared, including a parallel WaveNet model adaptively trained with the conventional distillation framework as our baseline and two proposed AGAN models with $\lambda=0.05$ and $\lambda=1.5$ respectively.
The ground-truth recordings are used in comparison and 63 professional English listeners participated in the listening test.
Since it is a neural vocoder, we focused on the fidelity (quality) of speech samples in our experiment.

The results of the subjective MOS evaluation are presented in Table~\ref{table:res}.
As we can see, our best model (AGAN with $\lambda=1.5$) performs better than the conventional adaptation approach (baseline). Absolute rise of 0.05 in MOS seems not large enough, but it’s worth noting that when MOS approaching natural speech, a tiny improvement (say, 0.01) represents notable improvements in some aspects of the human acoustical perception. The gap between the baseline parallel WaveNet model and ground truth natural speech is 0.1 in MOS. And our method further narrows this gap by half, making the speech generation achieving human level high fidelity\footnote{Audio samples can be found at \url{https://agan-demo.github.io/}.}.

We also investigated the importance of adversarial loss in AGAN by setting different values to \textbf{$\lambda$}. 
It can be easily analyzed from the results in Table~\ref{table:res} that the performance of AGAN significantly degraded when decreasing weight of adversarial loss, even worse than the baseline model.
And of course, more experiments are still needed to further demonstrate the relation between speech quality and adversarial loss.

\begin{figure}[t]

\begin{minipage}[b]{1\linewidth}
  \centering
  \centerline{\includegraphics[width=8cm, height=2cm]{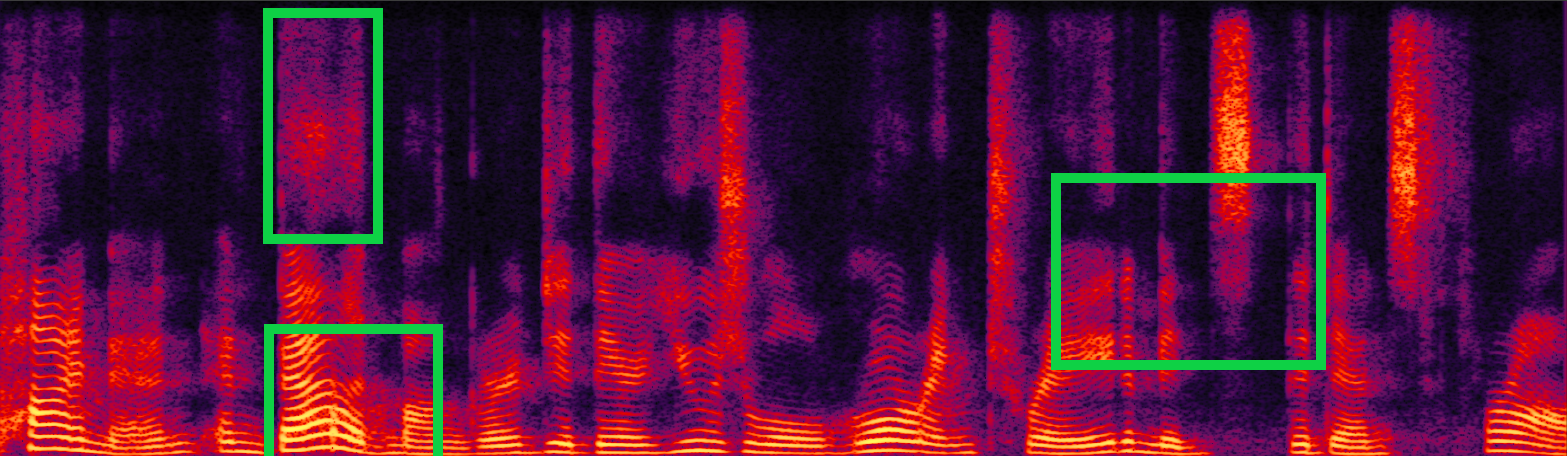}}
  \vspace{0.12cm}
  \centerline{(1-a) Parallel WaveNet}\medskip
\end{minipage}
\begin{minipage}[b]{1.0\linewidth}
  \centering
  \centerline{\includegraphics[width=8cm, height=2cm]{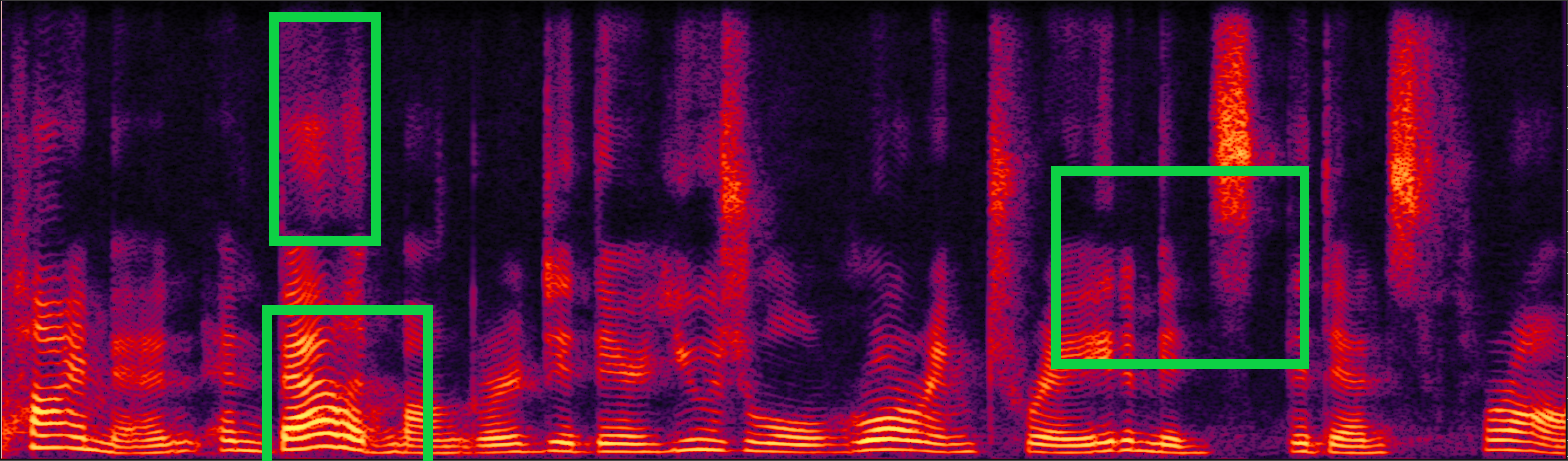}}
  \vspace{0.12cm}
  \centerline{(1-b) AGAN}\medskip
\end{minipage}
\begin{minipage}[b]{1.0\linewidth}
  \centering
  \centerline{\includegraphics[width=8cm, height=2cm]{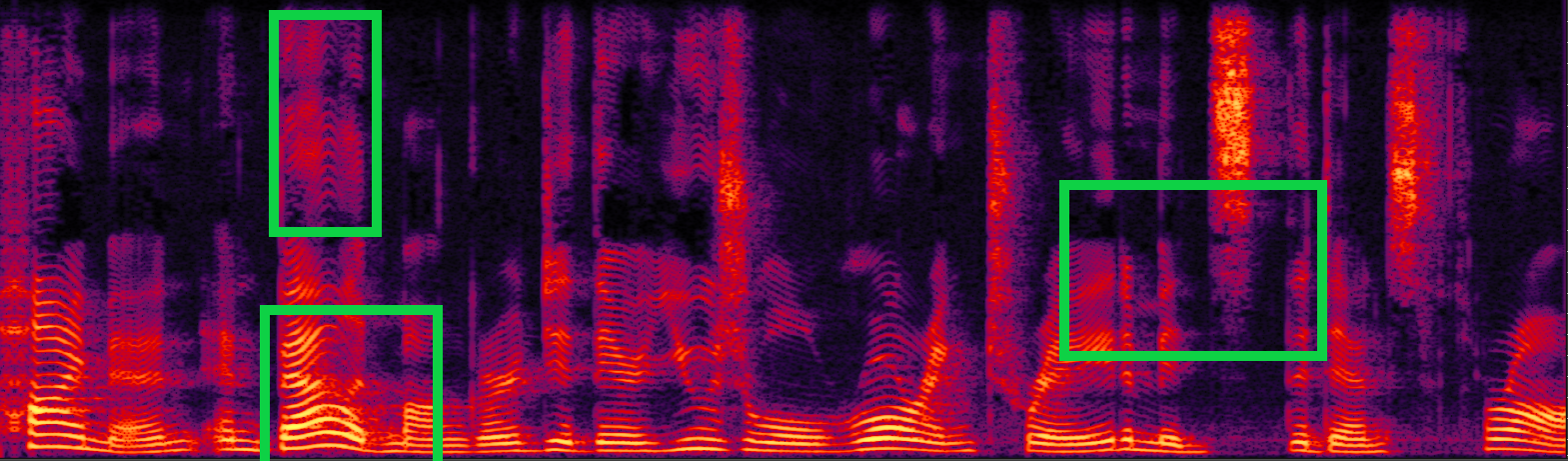}}
  \vspace{0.12cm}
  \centerline{(1-c) Ground-truth}\medskip
\end{minipage}

\begin{minipage}[H]{1.0\linewidth}
\centering
\includegraphics[width=2.5cm, height=2cm, width=.3\textwidth]{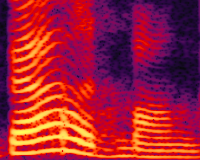}
\includegraphics[width=2.5cm, height=2cm, width=.3\textwidth]{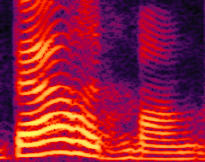}
\includegraphics[width=2.5cm, height=2cm, width=.3\textwidth]{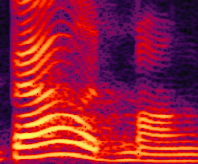}
\centerline{(1-d) Expanded Area}\medskip
\end{minipage}
\caption{STFT spectrograms of samples from parallel WaveNet, \textbf{AGAN (ours)} and Ground-Truth recording. Spectrograms of example 1. (1-d) subfigures are the expanded low frequency patterns of the lower left green-windowed areas(from left to right: parallel WaveNet, AGAN, Ground-truth). we can see that there exist some non-natural spectrum lines in parallel-WaveNet generated audio, while AGAN generated audio avoids such issue.}
\label{fig:res}

\end{figure}

\begin{figure}[t]

\begin{minipage}[b]{1\linewidth}
  \centering
  \centerline{\includegraphics[width=8cm, height=2cm]{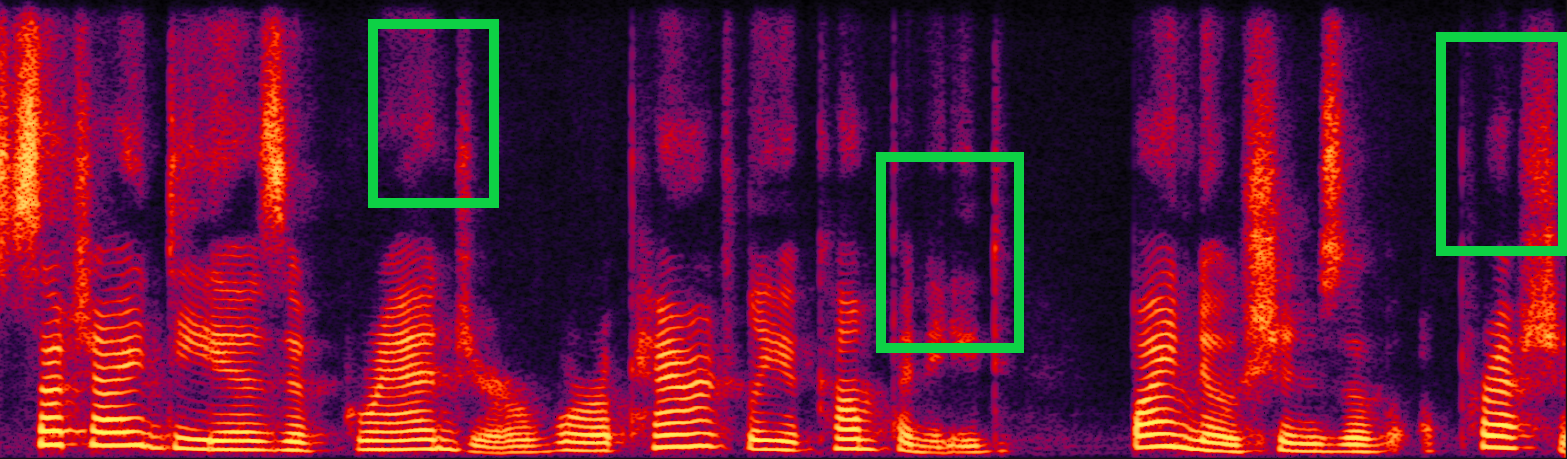}}
  \vspace{0.12cm}
  \centerline{(2-a) Parallel WaveNet}\medskip
\end{minipage}
\begin{minipage}[b]{1.0\linewidth}
  \centering
  \centerline{\includegraphics[width=8cm, height=2cm]{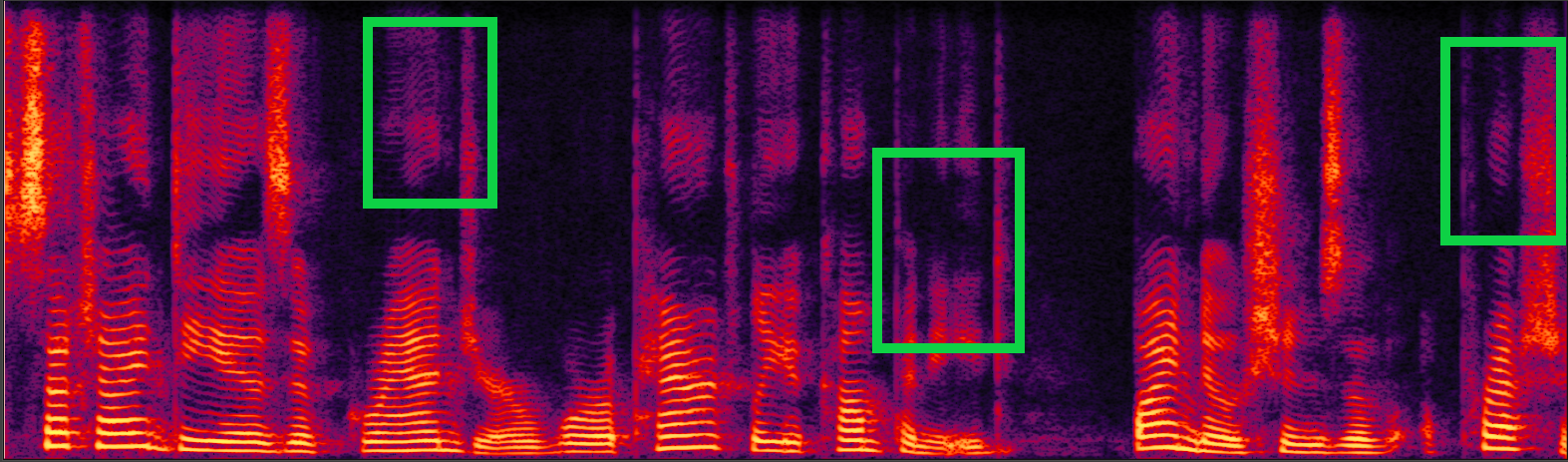}}
  \vspace{0.12cm}
  \centerline{(2-b) AGAN}\medskip
\end{minipage}
\begin{minipage}[b]{1.0\linewidth}
  \centering
  \centerline{\includegraphics[width=8cm, height=2cm]{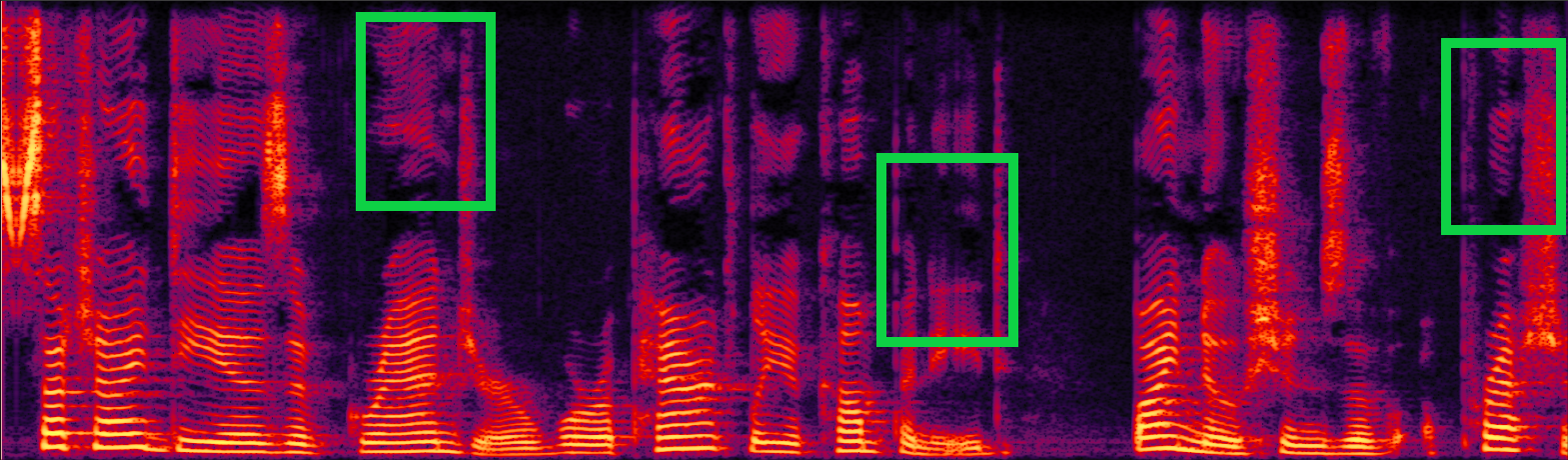}}
  \vspace{0.12cm}
  \centerline{(2-c) Ground-Truth}\medskip
\end{minipage}
\begin{minipage}[H]{1.0\linewidth}
\centering
\includegraphics[width=2.5cm, height=2cm, width=.3\textwidth]{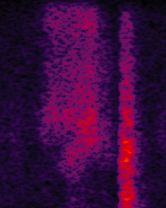}
\includegraphics[width=2.5cm, height=2cm, width=.3\textwidth]{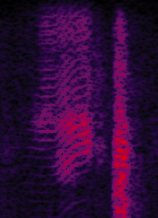}
\includegraphics[width=2.5cm, height=2cm, width=.3\textwidth]{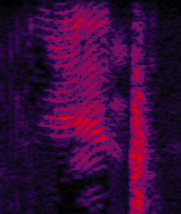}
\centerline{(2-d) Expanded Area}\medskip
\end{minipage}

\caption{STFT spectrograms of samples from parallel WaveNet, \textbf{AGAN (ours)} and Ground-Truth recording. Spectrograms of example 2. (2-d) subfigure expands the upper left green-windowed high frequency areas(from left to right: parallel WaveNet, AGAN, Ground-truth). From these figures we can tell that the resonance peaks are clear in AGAN output against the blurring spectrum in the baseline output ,which verifies that our AGAN method can generate more natural audios than parallel WaveNet}
\label{fig:res2}

\end{figure}

Apart from MOS evaluation, we also conducted a case study on the Mel-spectrograms of the models.
As shown in Fig.~\ref{fig:res} and Fig.~\ref{fig:res2}, we list two groups of audio samples which is consisting of ground-truth audio, PWN-generated audio and AGAN-generated audio spectrograms.  When comparing thoses Mel-spectrograms along time axis, it is clear that the proposed model can capture more detailed spectral information of the target speaker than parallel WaveNet model. Typical differences are marked with green windows in the figure. And in those areas, our AGAN model generated audios with resonance peaks that approach ground-truth audios better while parallel WaveNet generated lower quality spectragroms.
We can find obvious harmonic structures in the spectrograms of AGAN and ground truth generation results, but some of those details are missing in the parallel WaveNet baseline result.

\begin{table}[th]

\caption{Mean Opinion Score(MOS) with $95\%$ confidence intervals for different adaptation method.}
\label{table:res}
\begin{tabular}{cc}
\toprule
{\textbf{Method}} & {\textbf{Subjective 5-scale MOS}} \\
\midrule
Parallel WaveNet (baseline) & 4.53 $\pm$ 0.17 \\
AGAN ($\lambda$=0.05) & 4.50 $\pm$ 0.20 \\
\textbf{AGAN    ($\lambda$=1.50)} & \textbf{4.58 $\pm$ 0.16} \\
Ground-truth & 4.63 $\pm$ 0.14 \\
\bottomrule
\end{tabular}

\end{table}

\subsection{Adaptation cost of training}

Time consumption of adaptation training is vital for deployment in practical applications. Once the basic pre-trained model is obtained, the efficiency of adaptation determines the speed of training the entire vocoder system.
To prove the adaptation efficient of our proposed model,
we evaluate the training time of our method on a Nvidia Tesla P40 GPU.
It takes about 36 hours to complete an adaptation training for the baseline parallel WaveNet, which includes both teacher and student model.
The adaptation training of proposed AGAN on the same dataset takes about 12 hours, merely one third of the time that parallel WaveNet consumed.
This remarkable low time consumption is not only due to the efficiency of AGAN training, but also because that adaptation process is independent of a teacher model.
Although a discriminator is required in AGAN, its training is quite fast and stable.

\label{ssec:results of cost}

\section{Conclusions}
\label{sec:cons}

In this work, we propose  a speaker adaptation framework for parallel WaveNet vocoder based on GAN (AGAN). 
Comparing to conventional retrain-based model adaptation, 
AGAN performs more efficient adaptation on a relatively small amount of a new speaker data and generates speech with higher perceptual quality. 
And it provides an end to end adaptation method which is much faster than distilled framework such as parallel WaveNet. 
Our experiments indicates that the proposed method can further reduce the gap between speech samples from recording and proposed model. 

In addition, as a future work, it is straightforward that the proposed method can also be applied to optimize an IAF directly based parallel WaveNet model from scratch without the requirement of auto-regressive teacher model. 

\bibliographystyle{IEEEtran}

\bibliography{refs}

\end{document}